\documentclass[times, twocolumn]{aastex63}
\usepackage{gensymb}
\usepackage{amsmath}
\usepackage{graphicx}
\usepackage{textcomp}

\graphicspath{{./figures/}}

\begin{document}

\title{An Isolated White Dwarf with 317-Second Rotation and Magnetic Emission}

\author[0000-0003-1862-2951]{Joshua S. Reding}
\affil{University of North Carolina at Chapel Hill, Department of Physics and Astronomy, Chapel Hill, NC 27599, USA}

\author[0000-0001-5941-2286]{J. J. Hermes}
\affil{Boston University, Department of Astronomy, Boston, MA 02215, USA}

\author[0000-0002-0853-3464]{Z. Vanderbosch}
\affil{University of Texas at Austin, Department of Astronomy, Austin, TX 78712, USA}
\affil{McDonald Observatory, Fort Davis, TX 79734, USA}

\author[0000-0003-2852-268X]{E. Dennihy}
\affil{NSF's National Optical-Infrared Astronomy Research Laboratory \\ Gemini Observatory, Colina el Pino S/N, La Serena, Chile}

\author{B. C. Kaiser}
\affil{University of North Carolina at Chapel Hill, Department of Physics and Astronomy, Chapel Hill, NC 27599, USA}

\author{C. B. Mace}
\affil{University of North Carolina at Chapel Hill, Department of Physics and Astronomy, Chapel Hill, NC 27599, USA}

\author[0000-0002-1086-8685]{B. H. Dunlap}
\affil{University of Texas at Austin, Department of Astronomy, Austin, TX 78712, USA}

\author{J. C. Clemens}
\affil{University of North Carolina at Chapel Hill, Department of Physics and Astronomy, Chapel Hill, NC 27599, USA}

\begin{abstract}
We report the discovery of short-period photometric variability and modulated Zeeman-split hydrogen emission in SDSSJ125230.93-023417.72 (EPIC 228939929), a variable white dwarf star observed at long cadence in \textit{K2} Campaign 10. The behavior is associated with a magnetic ($B=5.0$ MG) spot on the stellar surface, making the $317.278$-second period a direct measurement of the stellar rotation rate. This object is therefore the fastest-rotating apparently isolated (without a stellar companion) white dwarf yet discovered, and the second found to exhibit chromospheric Balmer emission after GD 356, in which the emission has been attributed to a unipolar inductor mechanism driven by a possible rocky planet. We explore the properties and behavior of this object, and consider whether its evolution may hold implications for white dwarf mergers and their remnants.
\end{abstract}

\section{Introduction} \label{sec:intro}
\par During the near-decade of operation spanning both its original mission and \textit{K2}, the late \textit{Kepler} Space Telescope observed more than $1500$ white dwarfs, white dwarf candidates, and similar objects along the ecliptic plane (\citealp{Howell14}; \url{k2wd.org}). These data are a unique resource for exploring white dwarf variability; the quality and quantity of observations taken are unparalleled. There are a variety of isolated white dwarf variable types represented within the \textit{Kepler} sample: \citet{Vanderburg15} discovered transiting exoplanetary debris, \citet{Hermes17b} investigate pulsators, \citet{Hallakoun18} explore possible debris accretion, and \citet{Maoz15} and \citet{Hermes17a} discuss objects which present variation likely due to surface spots.
% (Transiting debris/planets)
\par This last category of spotted objects offers broad scientific value because surface spots provide a direct means of measuring stellar rotation rates. Work to date suggests that spots form in convective zones of magnetic white dwarfs \citep{Brinkworth13}, and may be present even at weak (kG) field strengths \citep{Hoard18}. Additionally, spots appear on white dwarfs both at low temperatures where atmospheres should be convective ($T_\mathrm{eff} < 12,\!000$K; \citealp{Tremblay11}), and at higher temperatures where atmospheres should be radiative \citep{Reding18}. This suggests that spots on isolated white dwarfs may appear either due to local suppression of convection, or due to other global effects such as surface distortion driven by magnetic pressure \citep{Fendt00}.
\par Regardless of the type of spot observed, the connection with magnetism raises questions about the progenitors of these white dwarfs. One potential explanation is that they are the remnants of white dwarf mergers, which are predicted to produce single stellar remnants \citep{Dan14}. This scenario would grant spotted white dwarfs a place of importance in the context of the type Ia supernova progenitor question, as double-degenerate mergers may contribute significantly to SN Ia rates \citep{Maoz18}.
\par The hot DQ (carbon-atmosphere) white dwarfs are thought to be an example of these merger remnants based on an inconsistency between kinematic and cooling ages \citep{Dunlap15}. \citet{Dufour13} also find that $\gtrsim\!70\%$ of these objects are magnetic, and some present variability at a single period (with harmonics) likely associated with rotation and surface spots. The periods of these variable objects are very short, on the order of $5$$-$$20$ minutes \citep{Williams16}, while most isolated white dwarfs rotate with periods of hours to days as determined from rotational broadening of spectral lines \citep{Berger05} and asteroseismology (\citealp{Kawaler15, Hermes17b}), suggesting that the merger process likely added significant angular momentum to the remnant.
\par In this paper, we present the discovery of SDSSJ125230.93-023417.72 (henceforth J1252), a white dwarf observed during \textit{K2} Campaign 10 that exhibits Zeeman-split Balmer emission and presents periodic modulation due to a magnetic spot on a rapid $317$-second timescale. This signal is observed in the \textit{Kepler} long-cadence data at the 11th super-Nyquist alias, which significantly diminishes the periodogram amplitude (Section~\ref{subsec:var}). We chronicle observations of this object from SDSS, \textit{Kepler}, SOAR, and McDonald Observatory in Section~\ref{sec:obs}, and follow with our analysis in Section~\ref{sec:nlss}. We then discuss broader implications suggested by this object in Section~\ref{sec:disc}, and present our conclusions in Section~\ref{sec:conc}.

\section{Observations} \label{sec:obs}
\par The Sloan Digital Sky Survey (SDSS) collected photometry of J1252 in Data Release 7 \citep{Girven11}, and \textit{Kepler} observed it in \textit{K2} Campaign 10 (EPIC 228939929) as a probable white dwarf candidate based on its blue color and high proper motion. Independently, the \textit{Galaxy Evolution Explorer} (\citealp{Martin05}; \textit{GALEX} 2414916899208956898) and the \textit{Wide-Field Infrared Survey Explorer} (\citealp{Wright10}; All\textit{WISE} J125230.98-023418.4) collected ultraviolet and infrared photometry, respectively (Table~\ref{table:phot} -- the All\textit{WISE} data has been converted into AB magnitude). \textit{Gaia} measured J1252 (\citealp{Perryman01}; DR2 3682469122383597056) to have a parallax of $\varpi = 12.94\pm0.11$ mas ($d=77.26\pm0.67$pc). This parallax is large and precise enough such that using $d = 1/\varpi$ should provide a sufficiently accurate estimate of the true distance \citep{Luri18}.

\subsection{K2 Photometry}
\par The \textit{Kepler K2} mission observed J1252 with long-cadence (29.4-min) exposures during Campaign 10, which ran from 6 July through 20 September 2016. \citet{Vanderburg14} processed this light curve using their \texttt{K2SFF} task.
\par
\begin{figure}[t!]
 \centerline{\includegraphics[width=0.975\columnwidth]{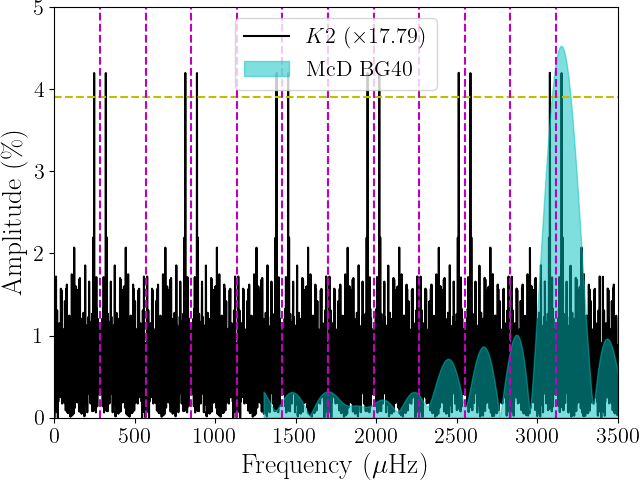}}
 \caption{Periodograms of J1252 from \textit{K2} (black) and McDonald Observatory BG40 (cyan) data; all aliases of the \textit{K2} long-cadence Nyquist frequency ($283.5\,\mu\mathrm{Hz}$) are marked as magenta dashed lines. We show a significance threshold of $5$ times the \textit{K2} periodogram average (yellow dashed line; \citealp{Baran15}). The \textit{K2} periodogram amplitude is scaled by a factor of $17.79$ to account for amplitude reduction over 11 Nyquist reflections \citep{Murphy14}. Our shorter ground-based observation cadence allowed us to identify the correct \textit{K2} periodogram feature at $3151.807\,\mu\mathrm{Hz}$, corresponding with a period of $317.278$ seconds.}
 \label{fig:periodogram}
\end{figure}

 J1252 unfortunately fell on CCD Module 4, which failed just over 7 days into Campaign 10. Despite this, the data collected were still sufficient to suggest variability with a semi-amplitude of at least $0.24\%$. However, we were not confident that we could identify the correct Nyquist alias from this periodogram alone (Fig.~\ref{fig:periodogram}). We therefore marked J1252 as a candidate for ground-based follow-up.

\subsection{SOAR/Goodman HTS Photometry}
\par The discovery of J1252's rapid variability resulted from three independent, yet serendipitous, hardware failures: the failure of the \textit{Kepler} reaction wheels, launching the \textit{K2} mission; the failure of CCD Module 4 during \textit{K2} Campaign 10, necessitating J1252's inclusion in our ground-based photometry plan; and the temporary failure of the Goodman High-Throughput Spectrograph's (HTS; \citealp{Clemens04}) slit-mask motor on our observing night. We observed J1252 on the $4.1$~m Southern Astrophysical Research (SOAR) Telescope on 24 April 2018 as part of a backup photometry plan, after this final failure left us unable to collect spectroscopy using Goodman. We collected a total of $2.9$ hours of integration time with $20$-second exposures using an \textit{S8612} broad-bandpass red-cutoff filter, with an overhead time between subsequent images of approximately $2.7$ seconds.
\par We bias- and flat field-corrected the data using median-combined master calibration images. We then used the Python package \texttt{photutils} \citep{PHOTUTILS} to centroid and fit a 2-dimensional gaussian profile to sources to determine a theoretically optimal circular aperture ($r_\mathrm{ap}$\texttildelow$1.6\sigma$; \citealp{Mighell99}). These apertures were typically $4\mathrm{-}5$ pixels in radius, and were surrounded by a background annulus starting 5 pixels from the terminus of the aperture and extending radially a further $10$ pixels. Using these parameters we performed photometry, de-trending for long-term effects using a quadratic polynomial and clipping the resultant light curve by $3\sigma$. We also applied a barycentric correction to the GPS-synced exposure midpoint timestamps using the Python package \texttt{astropy} (\citealp{ASTROPY1, ASTROPY2}). We isolated our analysis to the final \texttildelow$30$ minutes of this observation set, which were marginally less occulted with clouds.

{\renewcommand{\arraystretch}{1}
\begin{table}[t]
  \centering
  \caption{\textit{GALEX}, SDSS (PSF), and All\textit{WISE} survey photometry of J1252.\label{table:phot}}
  \begin{tabular*}{0.4\columnwidth}{l  c}
   \hline
   \hline
   Filter & AB Magnitude\\
   \hline
   \textit{FUV} & $22.68\pm0.22$\\
   \textit{NUV} & $19.232\pm0.024$\\
   \hline
   \textit{u} & $18.045\pm0.024$\\
   \textit{g} & $17.529\pm0.019$\\
   \textit{r} & $17.473\pm0.015$\\
   \textit{i} & $17.523\pm0.014$\\
   \textit{z} & $17.599\pm0.021$\\
   \hline
   \textit{W1} & $19.64\pm0.11$\\
   \textit{W2} & $20.00\pm0.31$\\
   \hline
   \end{tabular*}
\end{table}
}

{\renewcommand{\arraystretch}{}
\begin{table*}[t!]
  \centering
  \caption{Results of non-linear least squares fits of a sinusoidal signal to McDonald Observatory light curves from 2018 and 2019. The equation used is $A\sin[2\pi(t/P+\phi)]$, where the initial guess for $P$ is the period determined from the \textit{K2} data ($317.278\pm0.013$ s). We report the time of first photometric maximum in each filter for each set as determined from the phase term.\label{table:sinfit}}
  \begin{tabular*}{0.705\textwidth}{l | l l l l}
   \hline
   \hline
   Filter & A & P & $\phi$ & $t_\mathrm{0}$ (BJD\_TDB)\\
   \hline
   2018 & & & &\\
   BG40 & $4.56\pm0.14\%$ & $317.50\pm0.34\,\mathrm{s}$ & $0.9686\pm0.0099$ & $2458256.770258\pm0.000034$\\
   SDSS-\textit{g} & $4.89\pm0.19\%$ & $316.55\pm0.25\,\mathrm{s}$ & $0.215\pm0.012$ & $2458257.739876\pm0.000045$\\
   SDSS-\textit{i} & $2.22\pm0.20\%$ & $316.33\pm0.61\,\mathrm{s}$ & $0.238\pm0.030$ & $2458257.73994\pm0.00011$\\
   \hline
   2019 & & & &\\
   SDSS-\textit{g} & $5.19\pm0.21\%$ & $317.19\pm0.24\,\mathrm{s}$ & $0.534\pm0.013$ & $2458549.905111\pm0.000045$\\
   SDSS-\textit{i} & $2.22\pm0.23\%$ & $317.02\pm0.69\,\mathrm{s}$ & $0.584\pm0.034$ & $2458549.90507\pm0.00012$\\
   \hline
   \end{tabular*}
\end{table*}
}

\begin{figure}[t!]
   \centerline{\includegraphics[width=0.975\columnwidth]{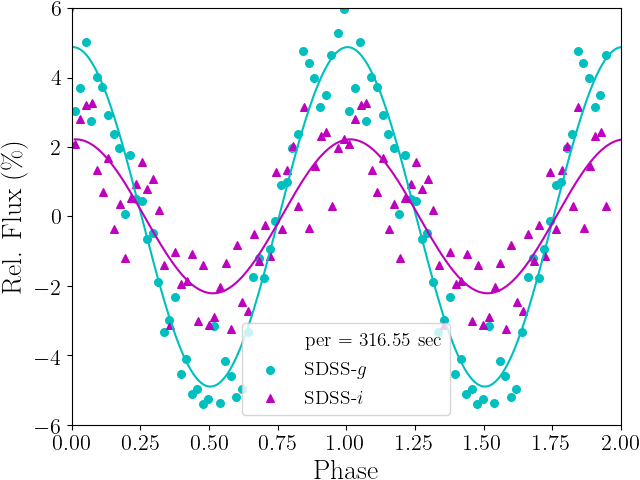}}
   \caption{Folded light curves of J1252 in SDSS-\textit{g} and \textit{i} filters from McDonald Observatory in 2018. We note a large ratio ($>\!2$) in amplitude between the two filters, and no apparent phase difference.}
   \label{fig:folded}
\end{figure}

\subsection{McDonald/ProEM Photometry}
\par Owing to the poor quality of the SOAR data, we collected additional photometric measurements of J1252 using the Otto Struve $2.1$ m Telescope at McDonald Observatory, Fort Davis, TX, on 18-19 May 2018 and on 7 March 2019. We used the ProEM 1024B frame-transfer CCD for data acquisition, which has been adapted to execute and repeat a user-defined sequence of filter changes. This enables collection of nearly simultaneous multi-band photometry by alternating the filters, making this system useful for exploring color-dependent behavior in variable objects.
\par On 18 May 2018, we observed J1252 for $1.43$ hours at a $10$-second cadence using a BG40 filter, which has a similar transmission profile to the S8612 filter used in our SOAR data. On 19 May, we swapped between SDSS-\textit{g} and \textit{i} filters in consecutive frames. This executed for $2.38$ hours at a $10$-second cadence, with a $3$-second blank frame between each exposure to allow the filter wheel to rotate. We then collected a final set with these same observation parameters on 7 March 2019 for $2.56$ hours.
\par We bias-, dark-, and flat field-corrected the McDonald data using standard calibration frames taken before each night of observations. Then, we performed circular aperture photometry to generate light curves using the Image Reduction and Analysis Facility (\texttt{IRAF}; \citealp{IRAF}) routine \texttt{CCD\_HSP} \citep{Kanaan02}. Treating each night and filter's light curve individually, we divided the light curves by the weighted mean of the comparison stars and normalized them by the average object flux.  We clipped out any extreme outliers and de-trended the light curves with quadratic polynomials to remove long-term trends.  We tested a range of aperture sizes and defined the optimal aperture based on the light curve with the minimum average point-to-point scatter. Lastly, we used the \texttt{WQED} software suite \citep{Thompson13} to apply a barycentric correction to the mid-exposure timestamps of each image.

\subsection{SOAR/Goodman HTS Spectroscopy}
We collected $2$ hours of time series spectroscopy of J1252 in $2$-minute exposures with SOAR and the Goodman HTS on 1 June 2019 using a $400$ line mm$^{-1}$ grating and $3.2''$ slit, corresponding to a slit width of $21\,$\AA. Our spectral resolution was therefore seeing-limited by the sky conditions at a FWHM of $7\,$\AA\ ($1''$). The average overhead for each acquisition was $5.45$ seconds. We bias-subtracted the data and trimmed the overscan regions, then completed reduction using a custom Python routine. We flux-calibrated the spectra using standard star EG 274, wavelength-calibrated using HgAr lamps, and applied a zero-point wavelength correction using sky lines from each exposure.

\section{Analysis}\label{sec:nlss}

\begin{figure*}[t!]
 \centerline{\includegraphics[width=0.975\textwidth]{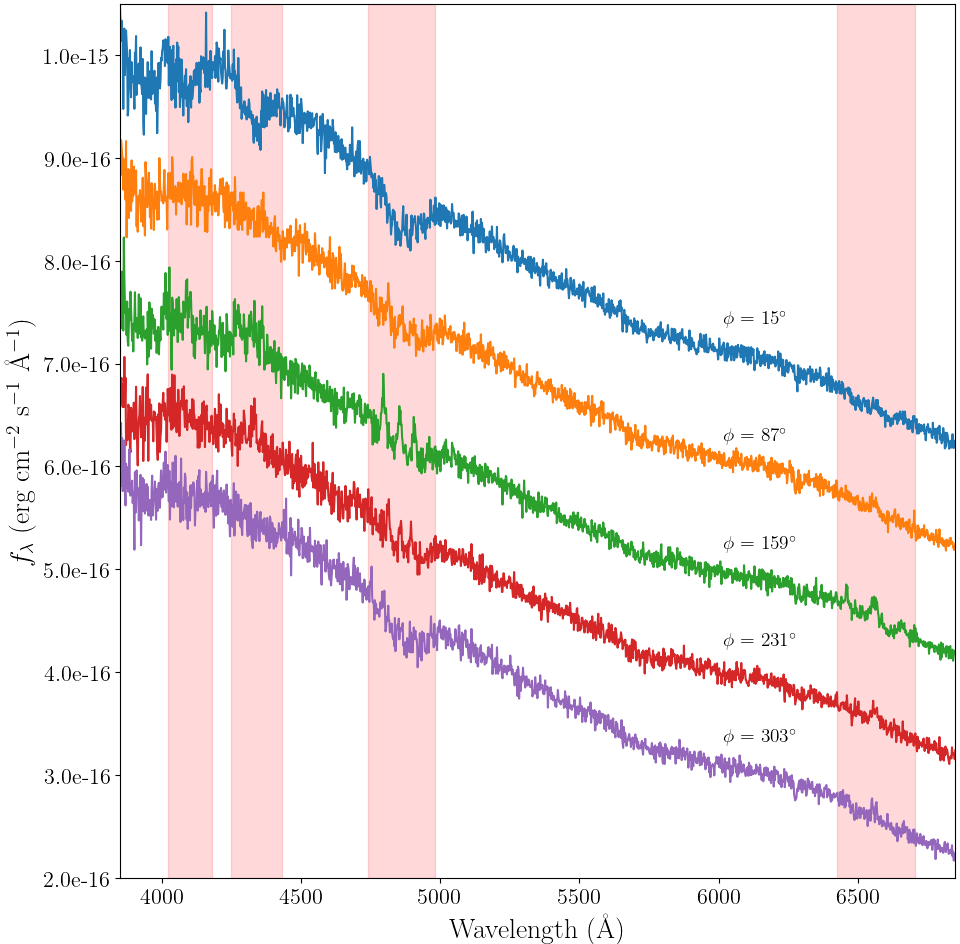}}
 \caption{Flux-calibrated and phase-binned SOAR spectra of J1252, offset from the bottom spectrum by increments of $2$e-$16$ erg cm$^{-2}$ s$^{-1}$ \AA$^{-1}$ for visibility. The absorption to emission transition is visible for Balmer features H$\alpha$-$\delta$ (shaded regions), and is most strongly evident at H$\beta$. Phase $\phi=0^\circ$ corresponds with the photometric maximum (emission minimum) in the SDSS-\textit{g} band. The phases reported are for the centers of each bin.}
 \label{fig:spec}
\end{figure*}

\subsection{Variability and Color Dependence}\label{subsec:var}
\par We computed a Lomb-Scargle periodogram from the \textit{K2} light curve using the \texttt{astropy.stats} Python package. In our periodogram, we ignore the first few integer harmonics of $47.2\,\mu\mathrm{Hz}$, as these result from the \textit{K2} drift correction thruster firing every $5.9$ hours \citep{Howell14}. We used the same tools to produce periodograms for our SOAR and McDonald data, whose rapid acquisition cadence allowed us to identify the correct \textit{K2} periodogram alias (Figure~\ref{fig:periodogram}).
\par After identifying the true alias, we performed a least squares fit of a sinusoid \big($A\sin[2\pi(t/P+\phi)]$\big) to the \textit{K2} light curve data with the software \texttt{Period04} \citep{Period04}, using the peak frequency of $3152\,\mu\mathrm{Hz}$ as an initial guess. Our best-fit value for the period of J1252 from \textit{K2} is $ 317.278\pm0.013$ seconds ($3151.807\pm0.013\,\mu\mathrm{Hz}$). When fitting to the ground-based data, we used this period as an initial guess, and performed non-linear least squares fits to determine if the period was consistent for all data sets (Table~\ref{table:sinfit}).

\par Because white dwarf rotation rates are typically on the order of hours to days (\citealp{Berger05, Hermes17b}), we were surprised to find that the period of this object was far shorter, falling at over 11 times the \textit{Kepler} long-cadence Nyquist frequency ($283.5\,\mu\mathrm{Hz}$). \citet{Murphy14} presents the function by which periodogram amplitude diminishes with super-Nyquist signals as $A_{\text{measured}}/A_{\text{intrinsic}} = |\text{sinc}(\pi t_\mathrm{exp}/P)|$, where $t_\mathrm{exp}$ is the instrument sampling rate, and $P$ is the object period. For J1252, the measured amplitude from the \textit{Kepler} long-cadence sampling rate is diminished by a factor of $17.79$ compared to the intrinsic. The inferred intrinsic amplitude from this is $4.20\%$, which is comparable to the $4.56\%$ amplitude from the McDonald BG40 data (Figure~\ref{fig:periodogram}). The remaining difference is attributable to the respective \textit{K2} and BG40 filter bandpass shapes and wavelength ranges.
\par We detected a significant difference in color amplitude, with ratio $g/i = 2.26\pm0.22$ in our multicolor light curves, but did not detect a significant phase shift between the two colors (Figure~\ref{fig:folded}). The individual fits also show evidence of period and phase modulations. The 2018 McDonald BG40 and SDSS filter sets were separated in acquisition by one day, but show a $2\!-\!3\sigma$ difference in the object period. In addition, there is significant power asymmetry in the lobes of the primary feature of the Fourier transform for the observational data, indicative of phase modulation. More data are needed to explore these phenomena further.

\begin{figure}[t!]
 \centerline{\includegraphics[width=1.0\columnwidth]{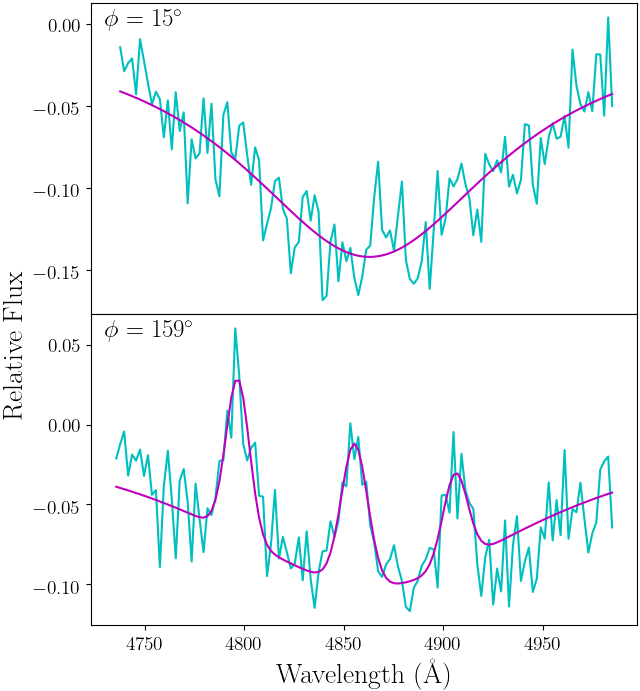}}
 \caption{H$\beta$ profiles from binned SOAR spectra (Figure~\ref{fig:spec}) at strongest observed absorption and emission phases, as measured from approximate photometric maximum ($\phi=0$). We determine feature locations from least squares fitting (solid line), and report results with corresponding magnetic field strengths in Table~\ref{table:fit}. The emission profile is consistent with an overall magnetic field strength of $B=5.0$ MG. Fluxes are relative to the continuum about the feature.}
 \label{fig:fit}
\end{figure}

{\renewcommand{\arraystretch}{1}
\begin{table}[t!]
  \centering
  \caption{Line locations and associated magnetic field strengths for H$\beta$ profile fit at emission minimum and maximum (Figure~\ref{fig:fit}). Field strength estimates are derived from \citet{Schimeczek14}.\label{table:fit}}
  \begin{tabular*}{0.67\columnwidth}{l c c}
   \hline
   \hline
   Line & Wavelength (\AA) & $B$ (MG)\\
   \hline
   Absorption & $4860.3\pm2.3$ & -\\
   Em. $\pi$ & $4855.3\pm0.8$ & $5.2\pm0.3$\\
   Em. $\sigma_-$ & $4797.0\pm0.8$ & $5.2\pm0.2$\\
   Em. $\sigma_+$ & $4906.2\pm1.0$ & $4.7\pm0.2$\\
   \hline
   \end{tabular*}
\end{table}
}

\subsection{Atmospheric Parameters and Magnetism}\label{subsec:nlss}
\par  We present the phase-binned $400$ l mm$^{-1}$ SOAR spectra of J1252 in Figure~\ref{fig:spec}. To produce these, we folded our individual spectra into five equally spaced phase bins, each covering one-fifth of the \textit{K2} variability period ($63.46$s; $12$ spectra per bin). We then averaged the spectra within each bin, using only those whose acquisition completely covered the phase window. We note that our acquisition times are larger than the phase windows, and so the resultant smearing between binned spectra complicates phase analysis. Also, given the apparent modulation in phase, we were unable to extrapolate phase associations for these window centers from previous data. Instead, we determined approximate phase values by convolving the binned spectra with the SDSS-\textit{g} filter transmission profile \citep{Doi10} and integrating to produce ``flux'' measurements, and fitting a sinusoid to these with period fixed to the \textit{K2} value. We defined the photometric maximum from this process as phase $\phi=0^\circ$.
\par The time-resolved $400$ l mm$^{-1}$ data revealed Balmer absorption features which deepen and diminish with variability phase, where the diminishment corresponds with the appearance of Zeeman-split emission. Variability and Zeeman-split Balmer emission has been seen before in GD 356 \citep{Greenstein85}, which was classified by \citet{Wesemael93} as a DAEH (hydrogen-dominated with magnetic emission) white dwarf. J1252 is thus the second star to bear this classification. To determine the magnetic field strength, we performed a least squares fit of the H$\beta$ profile at both absorption and emission maximum with the Python package \texttt{lmfit} \citep{Newville14}, using a Lorentzian profile for the wide absorption feature and Gaussian profiles for the emission peaks. The Zeeman emission is consistent with a surface-averaged magnetic field for J1252 of $5.0\pm0.1$ MG (Table~\ref{table:fit} and Figure~\ref{fig:fit}), as calculated using the H$\beta$ magnetic transitions catalogued in \citet{Schimeczek14}. This single value is a weighted average of the individual magnetic fields associated with each line shift, with weights determined by respective uncertainties. The spectral variability indicates that the magnetic spot is oriented along our line of sight at photometric minimum, which is the site of the strongest concentration of magnetic field lines. Therefore, this orientation provides the best measure of the polar magnetic field. Consequently, the absorption center in the phase of maximum absorption does not align with the central emission peak in the phase of maximum emission. This can be an effect of different hemisphere-averaged magnetic fields in the two phases.
\par Unfortunately, the effect of magnetism makes spectroscopic fitting of Balmer features to determine temperature and gravity for J1252 unreliable. We confirmed this by comparing our binned spectrum at the deepest absorption phase to model spectra from \citet{Koester10} (convolved with the night's seeing at $7$\AA), and found that even Balmer features corresponding with $\log$ g $=9.5$ were narrower than those we observed in J1252. Likewise, the expected equatorial rotation velocity of J1252 at \texttildelow$156 \mathrm{km}\,\mathrm{s}^{-1}$ ($R_*\!=\!7.89\!\times\!10^8 \mathrm{cm}$) would introduce Doppler broadening ($\Delta\lambda/\lambda = v/c$) on the order of $2.5$\AA at H$\beta$, a negligible effect compared to our observed feature widths.
\par Continuum fitting offers an alternative method to measure these parameters. \citet{GentileFusillo19} produced estimates for $\log$ g and ${T}_{\mathrm{eff}}$ for over \texttildelow$260,\!000$ white dwarfs, including J1252, by fitting \textit{Gaia} reddening-corrected absolute photometry to both hydrogen- and helium-model atmospheres (\citealp{Holberg06, Kowalski06, Tremblay11, Bergeron11}). Using hydrogen, the authors calculate best values for J1252 of ${T}_{\mathrm{eff}}=7642\pm99$ K and $\log$ g $=7.93\pm0.05$, which corresponds to a mass of $0.56\pm0.03$M$_\odot$.
\par To improve on the \citet{GentileFusillo19} results, we used the narrower-band SDSS survey photometry to produce more comprehensive photometric estimates of $\log$ g and ${T}_{\mathrm{eff}}$ for J1252. Because our time-series spectroscopy revealed Balmer absorption, we performed fitting of hydrogen-model atmospheres to the re-calibrated SDSS catalog PSF magnitudes \citep{Holberg06}. We interpolated the Bergeron DA models bicubically, and generated model photometry at each point for a grid of $\log$ g values spanning $7.50$-$8.50$ at a resolution of $0.01$, and ${T}_{\mathrm{eff}}$ spanning $7000$-$9000$ K every $1$ K. At each point in this grid, we evaluated the reduced $\chi^2$ of the observed photometry as fit to the generated model, and selected the ${T}_{\mathrm{eff}}$ and $\log$ g combination which produced a $\chi^2_\mathrm{red}$ closest to $1$. We noted that there is a large mismatch between the observed and predicted photometry for the near-UV range of \texttildelow$2000\!-\!3000\,$\AA, so we only perform fits using the SDSS-\textit{griz} points. We speculate on the origin of this deficit in Section~\ref{subsec:mag}.
\par The uncertainties in $\log$ g and ${T}_{\mathrm{eff}}$ include the photometric variability of the source, whose variability phase was unknown at the time of the SDSS acquisition. In order to account for this, we performed a Monte Carlo simulation. We perturbed each of the SDSS points $1000$ times according to a probability distribution function produced by convolving a Gaussian noise profile, generated from the mean errors on the SDSS PSF magnitudes, with an arcsine probability distribution expected from the assumed sinusoidal variability signal in each filter band, and performed model fitting with these new points. We did not correct the photometry for reddening, as J1252 is less than $100$pc from Earth, where reddening effects should be negligible \citep{GenestBeaulieu14}.
\par Our best estimates for J1252 are ${T}_{\mathrm{eff}}=8237\pm206$ K and $\log$ g $=8.09\pm0.05$, suggesting a mass of $0.65\pm0.03$M$_\odot$, which is higher than the \citet{GentileFusillo19} result and slightly above the average for white dwarfs ($0.62$M$_\odot$; \citealp{GenestBeaulieu19}).

\begin{figure}[t]
 \centerline{\includegraphics[width=1.0\columnwidth]{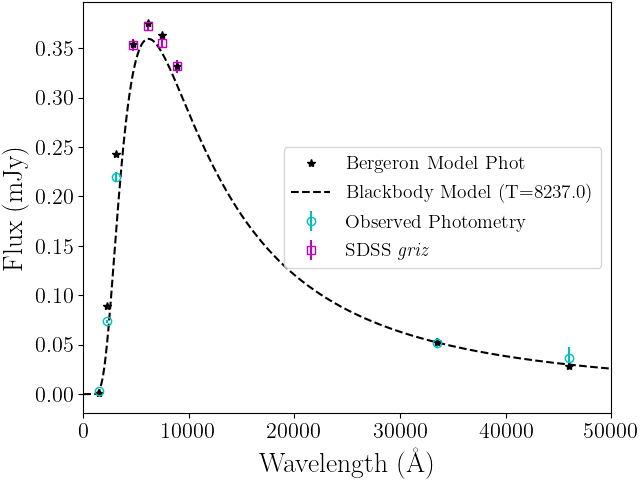}}
 \caption{Spectral energy distribution of J1252, calculated with survey photometry listed in Table~\ref{table:phot}. We also overplot a blackbody curve for the best stellar model, given our calculated $\log$ g $=8.09$ and ${T}_{\mathrm{eff}}=8237$ K determined by fitting SDSS-\textit{griz} points (marked as magenta squares) to the \citet{Bergeron01} DA models.}
 \label{fig:sed}
\end{figure}

\section{Discussion}\label{sec:disc}

\subsection{Variability Mechanism}\label{subsec:vard}
\par Our observations of periodic Zeeman-split Balmer emission confirm that J1252's photometric variations are consistent with a magnetic surface spot rotating in and out of view, with the photometric period corresponding to the star's rotation. The appearance and disappearance of the emission with rotation phase suggests that the excitation region is localized on the stellar surface.
\par
\begin{figure}[t]
 \centerline{\includegraphics[width=1.0\columnwidth]{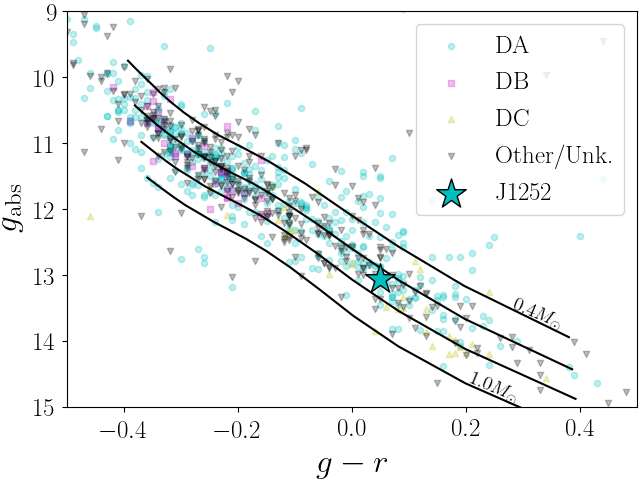}}
 \caption{SDSS color-magnitude diagram of white dwarfs observed in \textit{K2} (k2wd.org), with thick H layer mass tracks from \citet{Fontaine01}. J1252's photometry appears consistent with an apparently isolated (with no stellar companion) white dwarf of between $0.6$ and $0.7$M$_\odot$.}
 \label{fig:sdsscmd}
\end{figure}

 J1252 is also apparently isolated, with no detectable mass transfer from a stellar companion that might enhance its rotation rate and variability (Figure~\ref{fig:sdsscmd}). To place limits on the possibility of a stellar companion, we used the Database of Ultracool Parallaxes (\citealp{Dupuy12, Dupuy13, Liu16}) to produce spectral energy distributions for late-type stellar objects which had both \textit{WISE} photometry and known parallaxes. We then compared these SEDs to that of J1252 to determine the earliest companion spectral type which could be consistent with J1252's All\textit{WISE} photometry. While there was significant variation in \textit{W1} and \textit{W2} magnitudes for the brown dwarf regime, the \textit{WISE} photometry of the latest L-dwarfs (\texttildelow L9) consistently exceeded that of J1252 by approximately $3\sigma$ or more. We therefore may rule out a stellar or substellar companion earlier than spectral type T.
\par We also considered the possibility of an unseen white dwarf companion by fitting composite white dwarf models to the measured SED. These exercises show that we cannot formally rule out a very cool, very high-mass ($\lesssim\!5000$K, $\gtrsim\!1.2$M$_\odot$) white dwarf companion, or a binary of two white dwarfs with roughly equal temperature to J1252 whose combined mass is significantly super-Chandrasekhar ($\gtrsim\!1.9$M$_\odot$). While these contrived systems are consistent with our SED, we do not think them likely and they do not provide explanatory power for the observed magnetic emission. Absent any compelling evidence for binarity, we adopt the simplest fit to the SED of an apparently isolated white dwarf.
\par The only other isolated white dwarf observed to exhibit Zeeman-split hydrogen emission is GD 356 ($T_\mathrm{eff}=7510$ K, $\log g=8.14$), which has a $13$ MG polar magnetic field (\citealp{Greenstein85, Bergeron01, Wickramasinghe10}). It also shows low-amplitude (\texttildelow$0.2\%$) variability on $115$-min timescales, suggesting a surface spot which is never fully out of view due to the star's rotation axis orientation \citep{Brinkworth04}. Balmer emission is the only recognizable feature in GD 356's spectrum, though its SED fitting suggests a helium-dominated atmosphere \citep{Bergeron01}.
\par \citet{Li98} first suggested what has become the predominant model by which GD 356 exhibits emission. This model consists of a white dwarf with a dipole magnetic field which is orbited by a conducting body, such as a rocky planet. The orbit of this conducting body through the star's magnetosphere induces an electromotive force (and therefore a current) which is oriented along the magnetic field lines connecting the star and the companion. This type of system represents an example of a unipolar inductor. The closing of this ``circuit'' at either end heats both the orbiting object and the upper atmosphere of the star at its magnetic poles. If the companion is more conductive than the white dwarf (e.g. a rocky planet with no atmosphere, or a hotter white dwarf; \citealp{Li98}), ohmic dissipation of the energy from the system will occur preferentially in the primary white dwarf atmosphere, producing temperatures capable of exciting chromospheric gas into emission \citep{Wickramasinghe10}. This mechanism is known to be active in the Jupiter-Io system, where Io's orbit through Jupiter's magnetic field induces emission in both the radio and UV (\citealp{Goldreich69, Connerney93, Clarke02}).
\par \citet{Wickramasinghe10} suggest that a rocky planet driving the unipolar inductor in GD 356 potentially formed as the result of a double white dwarf merger, because it is unlikely that a rocky planet would survive the later stages of single-star evolution. The authors hypothesize that during a merger, one star may fully tidally disrupt and form a disk around the more massive companion. The disk may then cool from the exterior and begin to form dust and rocky material, and eventually planetary objects. This model has also been hypothesized for the origin of planets around millisecond pulsars \citep{Podsiadlowski91}.
\par A binary merger can produce a remnant with anomalously high-mass, which neither GD 356 nor J1252 appear to possess. However, \citet{Dan14} suggest that the merger of two low-mass white dwarfs ($\lesssim 0.4$M$_\odot$) can produce a single remnant with a mass near the white dwarf average. Furthermore, invoking a merger scenario potentially explains the appearance of high magnetic fields in GD 356 and J1252, which may be generated exclusively through mergers (\citealp{Tout08, Nordhaus11}). If the unipolar inductor model is also applicable to J1252, these two objects may represent a new class of white dwarf merger remnants --- magnetic emission line systems driven by a planetary dynamo mechanism.

\subsection{Near-UV Deficit and Color Amplitudes}\label{subsec:mag}
\par During our SED fitting process to determine estimates of J1252's atmospheric parameters, we noted that the near-UV filter fluxes (\textit{GALEX} NUV and SDSS-\textit{u}) fell well under the DA model atmosphere predictions. \citet{GentileFusillo18} identify a source for this discrepancy by using the weakly-magnetic WD2105-820 to show that white dwarf magnetism suppresses convection to the extent that radiative stellar models fit more accurately to observed spectra, even in temperature regimes where atmospheres should be convective ($T_\mathrm{eff} < 12,\!000$K; \citealp{Tremblay11}). This effect is particularly pronounced in the near-UV, where flux can be diminished on the order of $20\%$. Because J1252 is within this convective temperature regime and is strongly magnetic, the inclusion of convective mixing in the Bergeron models makes them a poor calibrator for J1252's near-UV flux.
\par We note as well that our SDSS-\textit{g}/\textit{i} amplitude ratio outscales more strongly magnetic white dwarfs with similar multicolor photometry \citep{Scholz18}. A potential explanation for both this and the UV deficit is the blanketing of flux in the near-UV due to metal pollution, which is then re-emitted in the optical. This ``fluorescence'' process \citep{Maoz15} is not an outlandish consequence of a system with potential rocky planetary debris; an estimated $30-50\%$ of white dwarfs experience metal pollution \citep{Koester14}. Furthermore, the amplitude of this variability may be accentuated by the channeling of material onto localized areas of the stellar surface, which results from magnetic activity. We speculate as well that the presence of strong emission at H$\beta$ also contributes to our large amplitude discrepancy.
\par In order to explore remaining questions on the presence of an unseen companion or near-UV activity, we encourage infrared and ultraviolet observations be taken for J1252. Furthermore, the unipolar inductor process may induce magnetic braking on the rotation of J1252, which would gradually slow the stellar rotation rate. Therefore, we also encourage ongoing monitoring of the variability period for J1252 in order to constrain this possibility, and to investigate the apparent phase modulation.

\section{Conclusions}\label{sec:conc}
\par SDSSJ125230.93-023417.72 ($0.65\pm0.03$M$_\odot$, ${T}_{\mathrm{eff}}=8237\pm206$ K), observed for 7 days during \textit{K2} Campaign 10, exhibits rapid photometric variability. Using follow-up spectroscopy and photometry from the $4.1$m SOAR Telescope and the McDonald Observatory $2.1$m Otto Struve Telescope, we were able to determine that the period of the variations is $317.278\pm0.013$ seconds. The spectrum (Figure~\ref{fig:spec}) is significantly variable, revealing a hydrogen (DA) white dwarf atmosphere at photometric maximum, and Zeeman-split Balmer emission at photometric minimum. Fits to the H$\beta$ emission suggest a magnetic field strength of $B = 5.0\pm0.1$ MG \citep{Schimeczek14}. J1252 therefore represents the second white dwarf with Balmer emission from a magnetic surface spot, after GD 356 \citep{Greenstein85}, and is the fastest-rotating apparently isolated (with no stellar companion) white dwarf yet discovered. J1252's spectral energy distribution also reveals a photometric deficit in the near-UV range of \texttildelow$2000\!-\!3000\,$\AA, which has been noted previously in the magnetic white dwarf WD2105-820 and others, and attributed to magnetic suppression of convection or metal line blanketing (\citealp{GentileFusillo18, Maoz15}).
\par The \textit{Gaia} astrometry and parallax ($\mu_\alpha=49.64$ mas yr$^{-1}$, $\mu_\delta=-39.95$ mas yr$^{-1}$, $\varpi = 12.94$ mas) of J1252 suggest a tangential velocity of $23.33$ km s$^{-1}$, which is consistent with the kinematics of \texttildelow$75\%$ of typical ($0.5 - 0.75$M$_\odot$) white dwarfs \citep{Wegg12}. If J1252 was produced through single stellar evolution, its effective temperature and surface gravity imply a \texttildelow $3$M$_\odot$ main sequence progenitor \citep{Cummings18}, and a cooling age of approximately $2$ Gyr, given a thick H layer \citep{Fontaine01}. \citet{Dunlap15} identified a discrepancy between stellar kinematics and implied cooling age for the hot DQ white dwarfs, and used this as a diagnostic to determine that these stars were merger remnants. Because the cooling age here appears consistent with the stellar kinematics, this analysis does not shed light on a merger origin for J1252. However, \citet{Dan14} theorize that a star of its mass can be the product of a double-degenerate merger.
\par J1252's analagous features to GD 356 suggest they share a similar variability mechanism. The predominant model for emission in GD 356 is a unipolar inductor, which involves a rocky planet inducing an electric current as it orbits through the host star's magnetosphere, thereby heating the star's upper atmosphere into emission \citep{Li98}. \citet{Wickramasinghe10} propose that such a planet may have formed after a double white dwarf merger. Furthermore, both stars exhibit magnetic fields which characterize them as high-field magnetic white dwarfs, which may exclusively be created in mergers (\citealp{Tout08, Nordhaus11}). There is also abundant evidence that mergers and mass transfer can speed up the rotation periods of white dwarfs, given that the fastest-rotating white dwarfs known are in cataclysmic variables (\citealp{Terada08, Mereghetti09}). These considerations suggest that J1252 is likely the result of a double-degenerate merger. J1252 and GD 356 may therefore represent a new class of white dwarf merger remnants --- magnetic white dwarfs with emission driven by a unipolar inductor mechanism.

\acknowledgements We would like to thank Mark Hollands for assistance with our Zeeman emission analysis.\\
Funding for this project was provided by NASA K2 Cycle 4 Grant NNX17AE92G. The \textit{K2} data may be obtained from the MAST archive at \dataset[doi:10.17909/T9RP4V]{https://dx.doi.org/10.17909/t9-1bev-a469}. This work is based on observations obtained at the Southern Astrophysical Research (SOAR) telescope, which is a joint project of the Minist\'{e}rio da Ci\^{e}ncia, Tecnologia, Inova\c{c}\~{o}es e Comunica\c{c}\~{o}es (MCTIC) do Brasil, the U.S. National Optical Astronomy Observatory (NOAO), the University of North Carolina at Chapel Hill (UNC), and Michigan State University (MSU). This paper includes data taken at McDonald Observatory of The University of Texas at Austin. Deepest thanks go to the McDonald observing support, especially John Kuehne, David Doss, and Coyne Gibson. This work has made use of data from the European Space Agency (ESA) mission {\it Gaia} (\url{https://www.cosmos.esa.int/gaia}), processed by the {\it Gaia} Data Processing and Analysis Consortium (DPAC, \url{https://www.cosmos.esa.int/web/gaia/dpac/consortium}). Funding for the DPAC has been provided by national institutions, in particular the institutions participating in the {\it Gaia} Multilateral Agreement. This work makes use of data products from the Wide-field Infrared Survey Explorer, which is a joint project of the University of California, Los Angeles, and the Jet Propulsion Laboratory/California Institute of Technology, funded by the National Aeronautics and Space Administration.
\software{\texttt{K2SFF} \citep{Vanderburg14}, \texttt{photutils} \citep{PHOTUTILS}, \texttt{astropy} (\citealp{ASTROPY1, ASTROPY2}), \texttt{IRAF} \citep{IRAF}, \texttt{CCD\_HSP} \citep{Kanaan02}, \texttt{WQED} \citep{Thompson13}, \texttt{Period04} \citep{Period04}, \texttt{lmfit} \citep{Newville14}}

\bibliography{/home/jsreding/Documents/UNC/Research/references.bib}

\end{document}